\documentclass[letter,11pt]{article}

\usepackage{amsmath,pslatex,color}

\usepackage{graphicx,color,amssymb,amsmath,subfigure}

\newcommand{\RR}{{\bf R}}
\newcommand{\RS}{{\bf S}}

\newtheorem{theorem}{Theorem}[section]
\newtheorem{lemma}[theorem]{Lemma}

\newcommand{\qed}{\nobreak \ifvmode \relax \else
      \ifdim\lastskip<1.5em \hskip-\lastskip
      \hskip1.5em plus0em minus0.5em \fi \nobreak
      \vrule height0.75em width0.5em depth0.25em\fi}

\begin{document}

\title{\bf Prediction of RNA-RNA interaction structure by centroids in the Boltzmann ensemble}

\author{Hamidreza Chitsaz\footnote{To whom correspondence should be addressed.}\\
Department of Computer Science and Engineering \\
University of California, San Diego\\
{\tt chitsaz@cs.ucsd.edu}}

\date{}

\maketitle

\begin{abstract}
\noindent New high-throughput sequencing technologies have made it possible to pursue
the advent of
 genome-wide transcriptomics. That progress combined with the recent discovery of regulatory non-coding RNAs (ncRNAs) has necessitated fast and accurate algorithms to predict RNA-RNA interaction probability and structure.
Although there are algorithms to predict minimum free energy interaction secondary structure for
two nucleic acids, little work has been done to exploit the information invested in the base pair probabilities to improve interaction structure prediction. 
In this paper, we present an algorithm to predict the Hamming centroid of
 the Boltzmann ensemble of interaction structures.
We also present an efficient algorithm to sample interaction structures from the ensemble. Our sampling algorithm uses a balanced scheme for traversing indices which improves the running time of the Ding-Lawrence sampling algorithm. The Ding-Lawrence sampling algorithm has $O(n^2m^2)$ time complexity whereas our algorithm has $O((n+m)^2\log(n+m))$ time complexity, in which $n$ and $m$ are the lengths of input strands.
We implemented our algorithm in a new version of {\tt piRNA} \cite{Chitsaz09} and compared our structure prediction results with competitors. Our centroid prediction
outperforms competitor minimum-free-energy prediction algorithms on average.
\end{abstract}

\section{Introduction}
The advent of
 genome-wide transcriptomics using high-throughput sequencing technologies
 and the recent discovery of regulatory non-coding RNAs (ncRNAs) have made it clear that RNA plays a
 large variety of important roles in living organisms
 that are more complex than being a mere intermediate in protein
 biosynthesis.
A large portion of these ncRNAs regulate gene expression
 post-transcriptionally through binding and forming base pairs (and establishing a joint structure) with 
 a target mRNA, like micro RNAs and small interfering RNAs (siRNAs)
 \cite{Bartel04, Hannon02, ZamHal05}, antisense RNAs~\cite{Bra02,WagFla02}
 or bacterial small regulatory RNAs (sRNAs) \cite{Gottesman05}. 
In addition, antisense
 oligonucleotides have been used as exogenous regulators of gene expression, usually to knock out
 genes for bacterial studies. Antisense technology is now commonly used as both a research tool and for therapeutic purposes. Synthetic nucleic acids have also been engineered to self assemble and interact in essentially nucleic acid machines \cite{SeeLuk05,See05,SimDit05,Ven07,Reif08}.

A key tool in all the above advances is an accurate tractable algorithm to predict the structure and base pairing probabilities between candidate regulatory ncRNAs and their potential targets. There are algorithms for predicting the most likely (the lowest total free energy) joint structure that can be formed
 by two interacting RNA strands \cite{AlkKarNadSahZha06}.
 Also, recently powerful algorithms for computing the partition function of interacting nucleic acid strands
 have been given (see our previous work \cite{Chitsaz09}, for example, or \cite{Huang09}). An important direction that has not been explored is to use the information invested in base pair probabilities to improve the accuracy of interaction structure prediction algorithms. In particular, Ding et al. take this promising direction but for prediction of the structure of a single nucleic acid strand \cite{Ding05}. In this paper, we aim to improve the accuracy of interaction structure prediction by centroids in the Boltzmann ensemble. 

In this paper, we present an algorithm to predict the Hamming centroid of
 an ensemble that is composed of the type of interactions that Alkan et al.~\cite{AlkKarNadSahZha06}
 considered.
We also present an efficient algorithm to sample interaction structures from the ensemble. Similar to the approach of \cite{Ding06}, sampled structures are clustered and the centroids of the clusters are considered as candidate structures. We believe success of such an approach critically depends on the clustering method, therefore, we leave sampling-clustering algorithms for a separate study.
 Our sampling algorithm uses a balanced scheme for traversing indices which improves the worst case running time complexity of the Ding-Lawrence sampling algorithm from $O(n^2m^2)$ to $O((n+m)^2\log(n+m))$, in which $n$ and $m$ are the lengths of input strands.
We implemented our algorithm in a new version of {\tt piRNA} \cite{Chitsaz09} and compared our structure prediction results with those of {\tt inteRNA}~\cite{AlkKarNadSahZha06} and the software of  Kato et al. \cite{Kato09}.
Our centroid prediction
outperforms competitor minimum-free-energy prediction algorithms in most of the experiments and on average.

\subsection*{Computational prediction of RNA secondary structure}
Several computational methods have emerged to study the
 secondary structure thermodynamics of a single nucleic acid
 strand. In the core of most methods lie a complete or variant of the 
Nearest Neighbor Thermodynamic energy model for 
a nucleic acid secondary structure \cite{MatTur99}. 
That model is widely considered the standard energy model. It is based on an (almost) log-linear Boltzmann probability
distribution founded on the assumption that stacking base
 pairs and loop entropies contribute additively to the free energy of a
 nucleic acid secondary structure. 
The standard energy model has been extended for pseudoknots 
and RNA-RNA interaction \cite{CaoChe06,Chitsaz09,DirPie03}. 
Exploiting the additivity of the energy, efficient divide and conquer algorithms for
 predicting the minimum free energy secondary structure \cite{Nus78,WatSmi78,Zuker81,RivEdd99} and 
 computing the partition function of a single strand \cite{Mcc90,DirPie03} have been developed.
 Also, Ding et al. give algorithms to predict the centroid of the Boltzmann ensemble and to sample structures from it \cite{Ding05,DinLaw03}. Ponty provides a new sampling algorithm, based on a balanced traversal of indices, whose worst case running time complexity is $O(n \log n)$ \cite{Ponty08}. Ponty's algorithm improves the running time complexity of the Ding-Lawrence algorithm, which is $O(n^2)$. 

\subsection*{Prediction of RNA-RNA interaction}
Initial methods to study the thermodynamics of multiple
 interacting strands concatenate input sequences in {\it silico} in some order and
 consider them as a single strand \cite{AndZha05, BerTaf06}.
Dirks et al. present a method, as a part of {\tt NUPack}, that computes
 the partition function for the whole ensemble of complex species
 carefully considering symmetry, sequence multiplicities, and special pseudoknots \cite{DirBoiSchWinPie07}. 
However, concatenating the sequences is not an accurate approach as even if pseudoknots are considered, some
 useful interactions are excluded while some physically impossible
 interactions are included. Some other methods simplify the problem by avoiding internal base-pairing in either strand,
 and compute the minimum free energy hybridization secondary structure   \cite{BerTaf06,DimZuk04,MarZuk08,RehSte04}. A third group predict the secondary structure of each individual RNA independently, and
 predict the (most likely) hybridization between the unpaired regions of two
 molecules \cite{BusRicBac08,Muckstein06,Walton02}.

In addition, a number of studies aim to compute
 the minimum free energy joint structure between two interacting strands
 under more complex structure and energy models.
Pervouchine devises a dynamic
 programming algorithm to maximize the number of base pairs among interacting strands \cite{Per04}.
Kato et al. propose a grammar based approach to RNA-RNA interaction prediction \cite{Kato09}.  
More generally, Alkan et al.~\cite{AlkKarNadSahZha06} study the interaction secondary structure prediction
 problem under three different models: 1) base pair counting, 2) stacked pair energy model,
 and 3) loop energy model. 
Alkan et al. prove that the general RNA-RNA interaction prediction under all three energy models is an 
 NP-hard problem. To reduce the complexity of the problem, 
 they suggest some natural constraints on the considered interaction secondary structures.
 These assumptions are satisfied by all
 examples of complex RNA-RNA interactions in the literature. 
The resulting algorithms efficiently compute the minimum free energy secondary
 structure among all possible joint secondary structures that do not contain
 (internal) pseudoknots, crossing interactions (i.e. external pseudoknots), and {\em zigzags} 
(please see section \ref{sec:pre} for the exact definition). In our previous work, we give a
dynamic programming algorithm
to compute the partition function over the ensemble of such interaction secondary structures \cite{Chitsaz09}.  

\section{Methods}
\label{sec:pre}
For the sake of completeness, we include here our notation and definitions given in \cite{Chitsaz09}.
Throughout this paper, we denote the two nucleic acid strands by $\RR$ and $\RS$. Strand $\RR$ is indexed from $1$ to $L_R$, and $\RS$ is indexed from $1$ to $L_S$ both in $5'$ to $3'$ direction. Note that the two strands interact in opposite directions, e.g. $\RR$ in $5' \rightarrow 3'$ with $\RS$ in $3' \leftarrow 5'$ direction. Each nucleotide is paired with at most one nucleotide in the same or the other strand. We refer to the $i^{th}$ nucleotide in $\RR$ and $\RS$ by $i_R$ and $i_S$ respectively. The subsequence from the $i^{th}$ nucleotide to the $j^{th}$ nucleotide in a strand is denoted by $[i, j]$.

An intramolecular base pair between the nucleotides $i$ and $j$ in a strand is called an {\it arc} and denoted by a bullet $i \bullet j$. 
An intermolecular base pair between the nucleotides $i_R$ and $i_S$ is called a {\it bond} and denoted by a circle $i_R \circ i_S$. 
An arc $i_R \bullet j_R$ \emph{covers} a bond $l_R \circ k_S$ if $i_R < l_R < j_R$. We call $i_R \bullet j_R$ an \emph{interaction arc} if there is a bond $l_R \circ k_S$ covered by $i_R \bullet j_R$. Assuming $i_R < j_R$, two bonds $i_R \circ i_S$ and $j_R \circ j_S$ are called \emph{crossing bonds} if $i_S < j_S$. 
An interaction arc $i_R \bullet j_R$ in a strand \emph{subsumes} a subsequence $[i_S, j_S]$ in the other strand if for all 
bonds $l_R \circ k_S$, if $i_S \leq k_S \leq j_S$ then $i_R < l_R < j_R $. 
Two interaction arcs $i_R \bullet j_R$ and $i_S \bullet j_S$ are part of a \emph{zigzag}, if neither $i_R \bullet j_R$ subsumes $[i_S, j_S]$ nor $i_S \bullet j_S$ subsumes $[i_R, j_R]$.

In this paper, we assume there are no pseudoknots in individual secondary structures of $\RR$ and $\RS$, and also there are no crossing bonds and zigzags between $\RR$ and $\RS$.

\subsection{Base pair probabilities and centroid prediction}
To estimate the centroid of the Boltzmann ensemble, it is sufficient to calculate the base pair probabilities
and select those base pairs whose probability is at least $0.5$. In this section, we describe how to calculate the base pair probabilities. Our algorithm for base pair probabilities is based on our dynamic programming algorithm for the interaction partition function {\tt piRNA} presented in \cite{Chitsaz09}.

Similar to {\tt piRNA}, our algorithm for base pair probabilities is also a dynamic programming algorithm that
computes two types of recursive quantities: 1) the probability of a subsequence $[i,j]$ in one strand, and 2) the probability of a joint subsequence pair $[i_R, j_R]$ and $[i_S, j_S]$. A \emph{region} is the domain over which a probability is computed. For the first type, region is $[i, j]$ and for the second type, region is $[i_R, j_R]\times [i_S, j_S]$. The \emph{length pair} of region $[i_R, j_R]\times [i_S, j_S]$ is $(l_R = j_R - i_R + 1,
 l_S = j_S - i_S + 1)$. Our algorithm starts with $(l_R=L_R, l_S=L_S)$ and considers all length pairs 
decrementally down to $(l_R = 1, l_S = 1)$. For a fixed length pair $(l_R, l_S)$, recursive quantities for all the regions $[i_R, i_R + l_R - 1]\times[i_S, i_S + l_S - 1]$ are computed.

\begin{figure}[h]
\begin{center}
\begin{picture}(0,0)%
\includegraphics{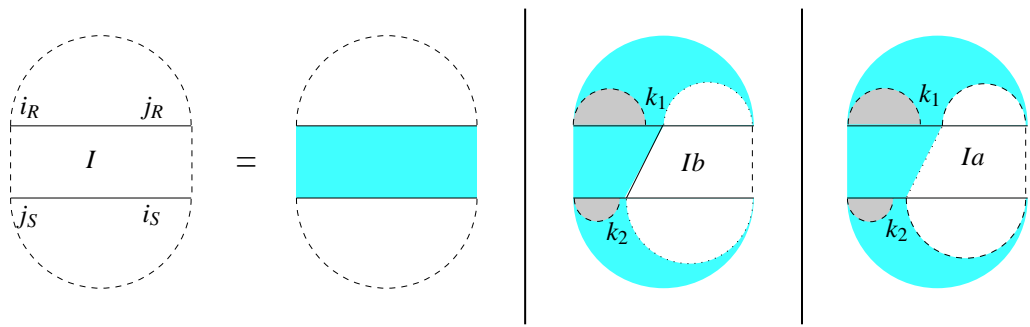}%
\end{picture}%
\setlength{\unitlength}{1657sp}%
\begingroup\makeatletter\ifx\SetFigFontNFSS\undefined%
\gdef\SetFigFontNFSS#1#2#3#4#5{%
  \reset@font\fontsize{#1}{#2pt}%
  \fontfamily{#3}\fontseries{#4}\fontshape{#5}%
  \selectfont}%
\fi\endgroup%
\begin{picture}(15254,4791)(-12486,-2839)
\put(-11339,-466){\makebox(0,0)[lb]{\smash{{\SetFigFontNFSS{10}{12.0}{\rmdefault}{\mddefault}{\updefault}{\color[rgb]{0,0,0}$I$}%
}}}}
\put(1756,-421){\makebox(0,0)[lb]{\smash{{\SetFigFontNFSS{10}{12.0}{\rmdefault}{\mddefault}{\updefault}{\color[rgb]{0,0,0}$Ia$}%
}}}}
\put(-2429,-511){\makebox(0,0)[lb]{\smash{{\SetFigFontNFSS{10}{12.0}{\rmdefault}{\mddefault}{\updefault}{\color[rgb]{0,0,0}$Ib$}%
}}}}
\put(-12329,344){\makebox(0,0)[lb]{\smash{{\SetFigFontNFSS{10}{12.0}{\rmdefault}{\mddefault}{\updefault}{\color[rgb]{0,0,0}$i_R$}%
}}}}
\put(631,-1501){\makebox(0,0)[lb]{\smash{{\SetFigFontNFSS{10}{12.0}{\rmdefault}{\mddefault}{\updefault}{\color[rgb]{0,0,0}$k_2$}%
}}}}
\put(1126,524){\makebox(0,0)[lb]{\smash{{\SetFigFontNFSS{10}{12.0}{\rmdefault}{\mddefault}{\updefault}{\color[rgb]{0,0,0}$k_1$}%
}}}}
\put(-3554,-1546){\makebox(0,0)[lb]{\smash{{\SetFigFontNFSS{10}{12.0}{\rmdefault}{\mddefault}{\updefault}{\color[rgb]{0,0,0}$k_2$}%
}}}}
\put(-2969,479){\makebox(0,0)[lb]{\smash{{\SetFigFontNFSS{10}{12.0}{\rmdefault}{\mddefault}{\updefault}{\color[rgb]{0,0,0}$k_1$}%
}}}}
\put(-10529,344){\makebox(0,0)[lb]{\smash{{\SetFigFontNFSS{10}{12.0}{\rmdefault}{\mddefault}{\updefault}{\color[rgb]{0,0,0}$j_R$}%
}}}}
\put(-10493,-1268){\makebox(0,0)[lb]{\smash{{\SetFigFontNFSS{10}{12.0}{\rmdefault}{\mddefault}{\updefault}{\color[rgb]{0,0,0}$i_S$}%
}}}}
\put(-12384,-1294){\makebox(0,0)[lb]{\smash{{\SetFigFontNFSS{10}{12.0}{\rmdefault}{\mddefault}{\updefault}{\color[rgb]{0,0,0}$j_S$}%
}}}}
\end{picture}%
\caption{\label{fig-i} Cases of the interaction partition function $Q^{I}_{i_R, j_R, i_S, j_S}$.
Figures \ref{fig-ib}, \ref{fig-ia} show the recursion for $Q^{Ib}$ and $Q^{Ia}$ where $b$ stands for bond and $a$ stands for arc.}
\end{center}
\end{figure}

For brevity, we present only two recursions and briefly describe how to derive the rest.
Let $P^{I}$, $P^{Ib}$, and $P^{Ia}$ be the probability of those substructures
that constitute respectively $Q^{I}$, $Q^{Ib}$, and $Q^{Ia}$ in {\tt piRNA} \cite{Chitsaz09}. Figure \ref{fig-i} shows the cases of $Q^{I}_{i_R, j_R, i_S, j_S}$ which is the interaction partition
function for the region $[i_R,j_R]\times[i_S,j_S]$. 
A horizontal line indicates the phosphate backbone, a solid curved line indicates an arc,
and a dashed curved line encloses a region and denotes its two terminal bases which may be paired or unpaired. Letter(s) within a region specify a recursive quantity. 
White regions are recursed over and blue regions indicate those portions of the secondary structure that are fixed at the current recursion level and contribute their energy to the partition function as defined by the energy model. A solid vertical line indicates a bond,
a dashed vertical line denotes two terminal bases of a region which may be base paired or unpaired, and a dotted 
vertical line denotes two terminal bases of a region which are assumed
to be unpaired. For the interaction partition functions, grey regions
indicate a reference to the partition functions for the single sequences.
 
The following equations precisely define the intended recursions: 
\begin{equation}\label{equ-i}
Q^{I}_{i_R, j_R, i_S, j_S}  = Q_{i_R, j_R}Q_{i_S, j_S} + 
 \sum_{i_R \leq k_1 < j_R \atop i_S < k_2 \leq j_S} Q_{i_R, k_1-1}Q_{k_2+1, j_S}Q^{Ib}_{k_1, j_R,i_S,k_2} +
 \sum_{i_R \leq k_1 < j_R \atop i_S < k_2 \leq j_S} Q_{i_R, k_1-1}Q_{k_2+1, j_S}Q^{Ia}_{k_1, j_R,i_S,k_2}, 
\end{equation}
 
\begin{equation}
Q^{Ib}_{i_R, j_R, i_S, j_S} = Q^{Ihh}_{i_R, j_R, i_S, j_S} + 
 \sum_{i_R < k_1 < j_R \atop i_S < k_2 < j_S} Q^{Ihb}_{i_R, k_1, k_2, j_S} Q^{Ib}_{k_1, j_R,i_S,k_2} +
\sum_{i_R < k_1 < j_R \atop i_S < k_2 < j_S} Q^{Ihh}_{i_R, k_1, k_2, j_S} Q^{Ia}_{k_1, j_R,i_S,k_2}, 
\label{equ-ib}
\end{equation}

\begin{figure}
\begin{center}
\begin{picture}(0,0)%
\includegraphics{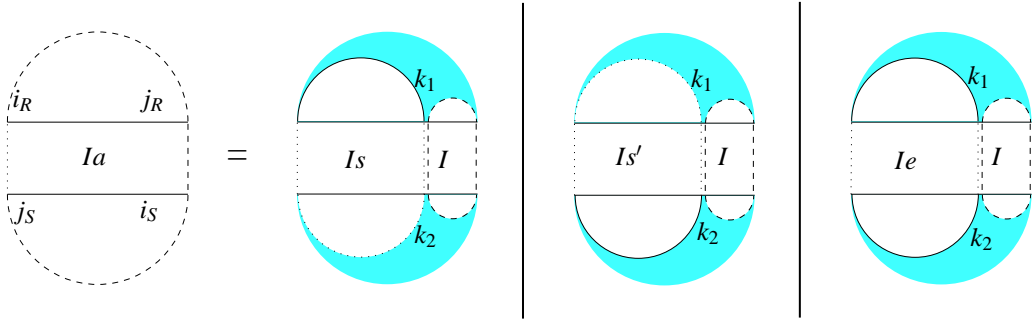}%
\end{picture}%
\setlength{\unitlength}{1657sp}%
\begingroup\makeatletter\ifx\SetFigFontNFSS\undefined%
\gdef\SetFigFontNFSS#1#2#3#4#5{%
  \reset@font\fontsize{#1}{#2pt}%
  \fontfamily{#3}\fontseries{#4}\fontshape{#5}%
  \selectfont}%
\fi\endgroup%
\begin{picture}(15359,4806)(8154,-12469)
\put(14626,-10186){\makebox(0,0)[lb]{\smash{{\SetFigFontNFSS{10}{12.0}{\rmdefault}{\mddefault}{\updefault}{\color[rgb]{0,0,0}$I$}%
}}}}
\put(18811,-10141){\makebox(0,0)[lb]{\smash{{\SetFigFontNFSS{10}{12.0}{\rmdefault}{\mddefault}{\updefault}{\color[rgb]{0,0,0}$I$}%
}}}}
\put(22906,-10141){\makebox(0,0)[lb]{\smash{{\SetFigFontNFSS{10}{12.0}{\rmdefault}{\mddefault}{\updefault}{\color[rgb]{0,0,0}$I$}%
}}}}
\put(13231,-10186){\makebox(0,0)[lb]{\smash{{\SetFigFontNFSS{10}{12.0}{\rmdefault}{\mddefault}{\updefault}{\color[rgb]{0,0,0}$Is$}%
}}}}
\put(21456,-10195){\makebox(0,0)[lb]{\smash{{\SetFigFontNFSS{10}{12.0}{\rmdefault}{\mddefault}{\updefault}{\color[rgb]{0,0,0}$Ie$}%
}}}}
\put(9301,-10141){\makebox(0,0)[lb]{\smash{{\SetFigFontNFSS{10}{12.0}{\rmdefault}{\mddefault}{\updefault}{\color[rgb]{0,0,0}$Ia$}%
}}}}
\put(17281,-10141){\makebox(0,0)[lb]{\smash{{\SetFigFontNFSS{10}{12.0}{\rmdefault}{\mddefault}{\updefault}{\color[rgb]{0,0,0}$Is'$}%
}}}}
\put(14266,-11311){\makebox(0,0)[lb]{\smash{{\SetFigFontNFSS{10}{12.0}{\rmdefault}{\mddefault}{\updefault}{\color[rgb]{0,0,0}$k_2$}%
}}}}
\put(14266,-8971){\makebox(0,0)[lb]{\smash{{\SetFigFontNFSS{10}{12.0}{\rmdefault}{\mddefault}{\updefault}{\color[rgb]{0,0,0}$k_1$}%
}}}}
\put(18406,-8971){\makebox(0,0)[lb]{\smash{{\SetFigFontNFSS{10}{12.0}{\rmdefault}{\mddefault}{\updefault}{\color[rgb]{0,0,0}$k_1$}%
}}}}
\put(18496,-11266){\makebox(0,0)[lb]{\smash{{\SetFigFontNFSS{10}{12.0}{\rmdefault}{\mddefault}{\updefault}{\color[rgb]{0,0,0}$k_2$}%
}}}}
\put(22591,-11356){\makebox(0,0)[lb]{\smash{{\SetFigFontNFSS{10}{12.0}{\rmdefault}{\mddefault}{\updefault}{\color[rgb]{0,0,0}$k_2$}%
}}}}
\put(22546,-8926){\makebox(0,0)[lb]{\smash{{\SetFigFontNFSS{10}{12.0}{\rmdefault}{\mddefault}{\updefault}{\color[rgb]{0,0,0}$k_1$}%
}}}}
\put(8273,-10943){\makebox(0,0)[lb]{\smash{{\SetFigFontNFSS{10}{12.0}{\rmdefault}{\mddefault}{\updefault}{\color[rgb]{0,0,0}$j_S$}%
}}}}
\put(8252,-9266){\makebox(0,0)[lb]{\smash{{\SetFigFontNFSS{10}{12.0}{\rmdefault}{\mddefault}{\updefault}{\color[rgb]{0,0,0}$i_R$}%
}}}}
\put(10158,-10924){\makebox(0,0)[lb]{\smash{{\SetFigFontNFSS{10}{12.0}{\rmdefault}{\mddefault}{\updefault}{\color[rgb]{0,0,0}$i_S$}%
}}}}
\put(10157,-9266){\makebox(0,0)[lb]{\smash{{\SetFigFontNFSS{10}{12.0}{\rmdefault}{\mddefault}{\updefault}{\color[rgb]{0,0,0}$j_R$}%
}}}}
\end{picture}%
\caption{\label{fig-ia} Cases of $Q^{Ia}_{i_R, j_R, i_S, j_S}$ for which we assume at least one of $i_R$ and $j_S$ is the end point of an interaction arc.}
\end{center}
\end{figure}

\begin{equation}
\begin{split}\label{equ-ia}
Q^{Ia}_{i_R, j_R, i_S, j_S} =& \sum_{i_R < k_1 \leq j_R \atop i_S \leq k_2 \leq j_S} Q^{Is}_{i_R, k_1,k_2,j_S} Q^{I}_{k_1+1,j_R,i_S, k_2-1} +  
 \sum_{i_R \leq k_1 \leq j_R \atop i_S < k_2 \leq j_S} Q^{Is'}_{i_R, k_1,k_2,j_S} Q^{I}_{k_1+1,j_R,i_S, k_2-1} + \\
 & \sum_{i_R < k_1 \leq j_R \atop i_S < k_2 \leq j_S} Q^{Ie}_{i_R, k_1,k_2,j_S} Q^{I}_{k_1+1,j_R,i_S, k_2-1}, 
\end{split}
\end{equation} 
in which $Q$, $Q^{Ihh}$, $Q^{Ihb}$, $Q^{Is}$, $Q^{Is'}$, and $Q^{Ie}$ are partition functions defined in \cite{Chitsaz09}. Note that among all partition function recursions given in \cite{Chitsaz09}, $Q^{I}$ appears on the right hand side of only (\ref{equ-ia}). Therefore,
\begin{equation}
P^{I}_{i_R, j_R, i_S, j_S} = \sum_{1 \leq k_1 < i_R \atop j_S < k_2 \leq L_S} 
P^{Ia}_{k_1, j_R, i_S, k_2} \frac{(Q^{Is}_{k_1, i_R, j_S, k_2} + Q^{Is'}_{k_1, i_R, j_S, k_2} + Q^{Ie}_{k_1, i_R, j_S, k_2}) Q^{I}_{i_R, j_R, i_S, j_S}}
{Q^{Ia}_{k_1, j_R, i_S, k_2}},
\label{equ:p}
\end{equation}
with $P^{I}_{1, L_R, 1, L_S} = 1$ as the initial condition. Also note that $Q^{Ia}$ appears on the right hand side of only (\ref{equ-i}) and (\ref{equ-ib}), hence,
\begin{equation}
P^{Ia}_{i_R, j_R, i_S, j_S} = \sum_{1 \leq k_1 \leq i_R \atop j_S \leq k_2 \leq L_S} 
P^{I}_{k_1, j_R, i_S, k_2}\frac{Q_{k_1, i_R-1}Q_{j_S+1, k_2} Q^{Ia}_{i_R, j_R, i_S, j_S}}{Q^{I}_{k_1, j_R, i_S, k_2}} + 
\sum_{1 \leq k_1 < i_R \atop j_S \leq k_2 \leq L_S} 
P^{Ib}_{k_1, j_R, i_S, k_2}\frac{Q^{Ihh}_{k_1, i_R, j_S, k_2} Q^{Ia}_{i_R, j_R, i_S, j_S}}{Q^{Ib}_{k_1, j_R, i_S, k_2}}.
\label{equ-pia}
\end{equation} 
Using the same technique, by considering the contribution of each right hand side term that contains the target partition function, all the probability recursions are derived; see \cite{DirPie04} for more details of the technique. Finally, base pair probabilities are

\begin{figure}[h]
\begin{center}
\begin{picture}(0,0)%
\includegraphics{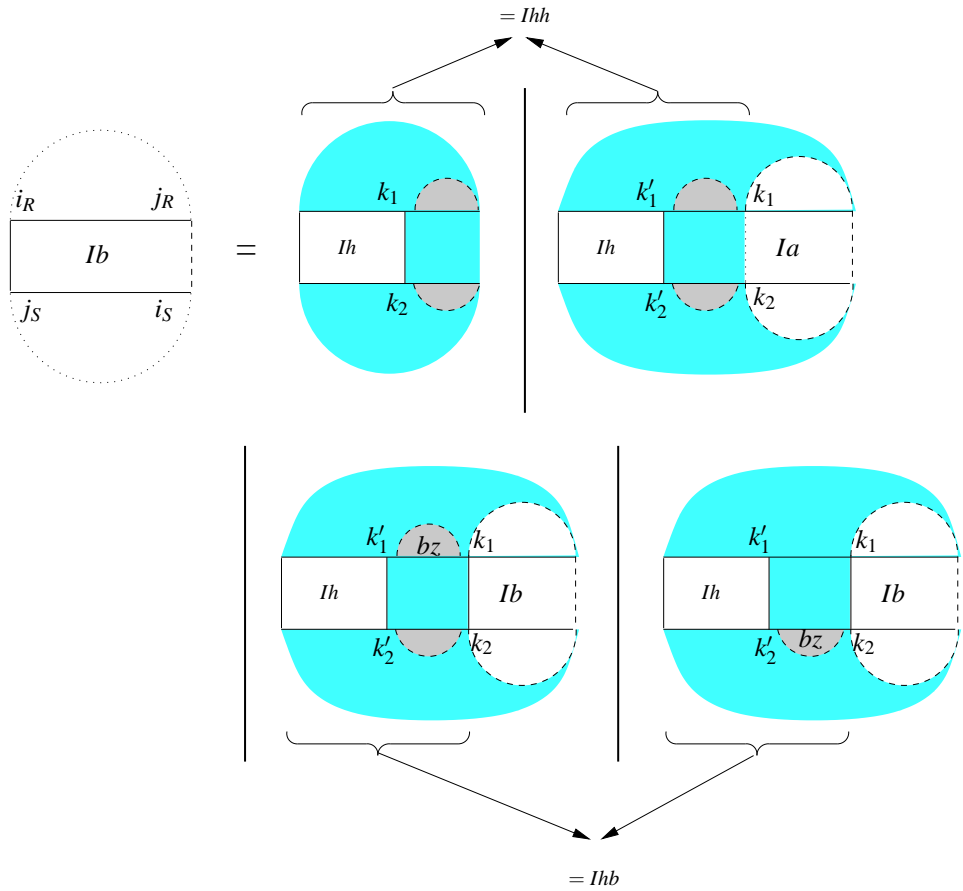}%
\end{picture}%
\setlength{\unitlength}{1657sp}%
\begingroup\makeatletter\ifx\SetFigFontNFSS\undefined%
\gdef\SetFigFontNFSS#1#2#3#4#5{%
  \reset@font\fontsize{#1}{#2pt}%
  \fontfamily{#3}\fontseries{#4}\fontshape{#5}%
  \selectfont}%
\fi\endgroup%
\begin{picture}(14243,13390)(-12486,-5462)
\put(-11339,3989){\makebox(0,0)[lb]{\smash{{\SetFigFontNFSS{10}{12.0}{\rmdefault}{\mddefault}{\updefault}{\color[rgb]{0,0,0}$Ib$}%
}}}}
\put(-7564,4124){\makebox(0,0)[lb]{\smash{{\SetFigFontNFSS{8}{9.6}{\rmdefault}{\mddefault}{\updefault}{\color[rgb]{0,0,0}$Ih$}%
}}}}
\put(-4769,7589){\makebox(0,0)[b]{\smash{{\SetFigFontNFSS{10}{12.0}{\sfdefault}{\mddefault}{\updefault}{\color[rgb]{0,0,0}\fontsize{8}{9}\selectfont $=$ $Ihh$}%
}}}}
\put(-7834,-1051){\makebox(0,0)[lb]{\smash{{\SetFigFontNFSS{8}{9.6}{\rmdefault}{\mddefault}{\updefault}{\color[rgb]{0,0,0}$Ih$}%
}}}}
\put(-5149,-1091){\makebox(0,0)[lb]{\smash{{\SetFigFontNFSS{10}{12.0}{\rmdefault}{\mddefault}{\updefault}{\color[rgb]{0,0,0}$Ib$}%
}}}}
\put(-2119,-1051){\makebox(0,0)[lb]{\smash{{\SetFigFontNFSS{8}{9.6}{\rmdefault}{\mddefault}{\updefault}{\color[rgb]{0,0,0}$Ih$}%
}}}}
\put(566,-1091){\makebox(0,0)[lb]{\smash{{\SetFigFontNFSS{10}{12.0}{\rmdefault}{\mddefault}{\updefault}{\color[rgb]{0,0,0}$Ib$}%
}}}}
\put(-3694,4124){\makebox(0,0)[lb]{\smash{{\SetFigFontNFSS{8}{9.6}{\rmdefault}{\mddefault}{\updefault}{\color[rgb]{0,0,0}$Ih$}%
}}}}
\put(-1009,4084){\makebox(0,0)[lb]{\smash{{\SetFigFontNFSS{10}{12.0}{\rmdefault}{\mddefault}{\updefault}{\color[rgb]{0,0,0}$Ia$}%
}}}}
\put(-3734,-5326){\makebox(0,0)[b]{\smash{{\SetFigFontNFSS{10}{12.0}{\sfdefault}{\mddefault}{\updefault}{\color[rgb]{0,0,0} \fontsize{8}{9}\selectfont $=$ $Ihb$}%
}}}}
\put(-12329,3179){\makebox(0,0)[lb]{\smash{{\SetFigFontNFSS{10}{12.0}{\rmdefault}{\mddefault}{\updefault}{\color[rgb]{0,0,0}$j_S$}%
}}}}
\put(-10304,3179){\makebox(0,0)[lb]{\smash{{\SetFigFontNFSS{10}{12.0}{\rmdefault}{\mddefault}{\updefault}{\color[rgb]{0,0,0}$i_S$}%
}}}}
\put(-6974,4934){\makebox(0,0)[lb]{\smash{{\SetFigFontNFSS{10}{12.0}{\rmdefault}{\mddefault}{\updefault}{\color[rgb]{0,0,0}$k_1$}%
}}}}
\put(-6794,-241){\makebox(0,0)[rb]{\smash{{\SetFigFontNFSS{10}{12.0}{\rmdefault}{\mddefault}{\updefault}{\color[rgb]{0,0,0}$k'_1$}%
}}}}
\put(-5534,-279){\makebox(0,0)[lb]{\smash{{\SetFigFontNFSS{10}{12.0}{\rmdefault}{\mddefault}{\updefault}{\color[rgb]{0,0,0}$k_1$}%
}}}}
\put(-1079,-241){\makebox(0,0)[rb]{\smash{{\SetFigFontNFSS{10}{12.0}{\rmdefault}{\mddefault}{\updefault}{\color[rgb]{0,0,0}$k'_1$}%
}}}}
\put(181,-279){\makebox(0,0)[lb]{\smash{{\SetFigFontNFSS{10}{12.0}{\rmdefault}{\mddefault}{\updefault}{\color[rgb]{0,0,0}$k_1$}%
}}}}
\put(-3104,4934){\makebox(0,0)[lb]{\smash{{\SetFigFontNFSS{10}{12.0}{\rmdefault}{\mddefault}{\updefault}{\color[rgb]{0,0,0}$k'_1$}%
}}}}
\put(-1349,4896){\makebox(0,0)[lb]{\smash{{\SetFigFontNFSS{10}{12.0}{\rmdefault}{\mddefault}{\updefault}{\color[rgb]{0,0,0}$k_1$}%
}}}}
\put(-12391,4820){\makebox(0,0)[lb]{\smash{{\SetFigFontNFSS{10}{12.0}{\rmdefault}{\mddefault}{\updefault}{\color[rgb]{0,0,0}$i_R$}%
}}}}
\put(-10366,4861){\makebox(0,0)[lb]{\smash{{\SetFigFontNFSS{10}{12.0}{\rmdefault}{\mddefault}{\updefault}{\color[rgb]{0,0,0}$j_R$}%
}}}}
\put(-6839,3269){\makebox(0,0)[lb]{\smash{{\SetFigFontNFSS{10}{12.0}{\rmdefault}{\mddefault}{\updefault}{\color[rgb]{0,0,0}$k_2$}%
}}}}
\put(-6704,-1906){\makebox(0,0)[rb]{\smash{{\SetFigFontNFSS{10}{12.0}{\rmdefault}{\mddefault}{\updefault}{\color[rgb]{0,0,0}$k'_2$}%
}}}}
\put(-5579,-1816){\makebox(0,0)[lb]{\smash{{\SetFigFontNFSS{10}{12.0}{\rmdefault}{\mddefault}{\updefault}{\color[rgb]{0,0,0}$k_2$}%
}}}}
\put(-989,-1906){\makebox(0,0)[rb]{\smash{{\SetFigFontNFSS{10}{12.0}{\rmdefault}{\mddefault}{\updefault}{\color[rgb]{0,0,0}$k'_2$}%
}}}}
\put(136,-1816){\makebox(0,0)[lb]{\smash{{\SetFigFontNFSS{10}{12.0}{\rmdefault}{\mddefault}{\updefault}{\color[rgb]{0,0,0}$k_2$}%
}}}}
\put(-2969,3269){\makebox(0,0)[lb]{\smash{{\SetFigFontNFSS{10}{12.0}{\rmdefault}{\mddefault}{\updefault}{\color[rgb]{0,0,0}$k'_2$}%
}}}}
\put(-1349,3359){\makebox(0,0)[lb]{\smash{{\SetFigFontNFSS{10}{12.0}{\rmdefault}{\mddefault}{\updefault}{\color[rgb]{0,0,0}$k_2$}%
}}}}
\put(-6209,-376){\makebox(0,0)[b]{\smash{{\SetFigFontNFSS{10}{12.0}{\rmdefault}{\mddefault}{\updefault}{\color[rgb]{0,0,0}$bz$}%
}}}}
\put(-494,-1771){\makebox(0,0)[b]{\smash{{\SetFigFontNFSS{10}{12.0}{\rmdefault}{\mddefault}{\updefault}{\color[rgb]{0,0,0}$bz$}%
}}}}
\end{picture}%
\caption{\label{fig-ib} Recursion for $Q^{Ib}_{i_R, j_R, i_S, j_S}$
  assuming $i_R \circ j_S$ is a bond.}
\end{center}
\end{figure}
\begin{figure}[h]
\begin{center}
\begin{picture}(0,0)%
\includegraphics{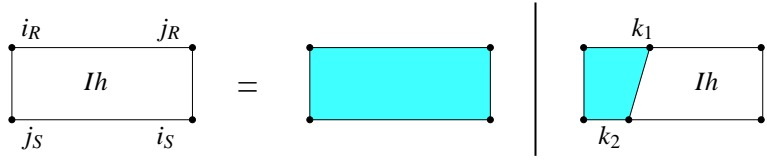}%
\end{picture}%
\setlength{\unitlength}{1657sp}%
\begingroup\makeatletter\ifx\SetFigFontNFSS\undefined%
\gdef\SetFigFontNFSS#1#2#3#4#5{%
  \reset@font\fontsize{#1}{#2pt}%
  \fontfamily{#3}\fontseries{#4}\fontshape{#5}%
  \selectfont}%
\fi\endgroup%
\begin{picture}(11325,2361)(-12517,-1489)
\put(-2249,-466){\makebox(0,0)[lb]{\smash{{\SetFigFontNFSS{10}{12.0}{\rmdefault}{\mddefault}{\updefault}{\color[rgb]{0,0,0}$Ih$}%
}}}}
\put(-11384,-466){\makebox(0,0)[lb]{\smash{{\SetFigFontNFSS{10}{12.0}{\rmdefault}{\mddefault}{\updefault}{\color[rgb]{0,0,0}$Ih$}%
}}}}
\put(-10304,344){\makebox(0,0)[lb]{\smash{{\SetFigFontNFSS{10}{12.0}{\rmdefault}{\mddefault}{\updefault}{\color[rgb]{0,0,0}$j_R$}%
}}}}
\put(-12329,344){\makebox(0,0)[lb]{\smash{{\SetFigFontNFSS{10}{12.0}{\rmdefault}{\mddefault}{\updefault}{\color[rgb]{0,0,0}$i_R$}%
}}}}
\put(-12329,-1276){\makebox(0,0)[lb]{\smash{{\SetFigFontNFSS{10}{12.0}{\rmdefault}{\mddefault}{\updefault}{\color[rgb]{0,0,0}$j_S$}%
}}}}
\put(-10304,-1276){\makebox(0,0)[lb]{\smash{{\SetFigFontNFSS{10}{12.0}{\rmdefault}{\mddefault}{\updefault}{\color[rgb]{0,0,0}$i_S$}%
}}}}
\put(-3194,344){\makebox(0,0)[lb]{\smash{{\SetFigFontNFSS{10}{12.0}{\rmdefault}{\mddefault}{\updefault}{\color[rgb]{0,0,0}$k_1$}%
}}}}
\put(-3689,-1276){\makebox(0,0)[lb]{\smash{{\SetFigFontNFSS{10}{12.0}{\rmdefault}{\mddefault}{\updefault}{\color[rgb]{0,0,0}$k_2$}%
}}}}
\end{picture}%
\caption{\label{fig-ih} Cases of $Q^{Ih}_{i_R, j_R, i_S, j_S}$ the
  interaction partition function for a single hybrid component.}
\end{center}
\end{figure}


\begin{align}
P(i_R, j_R) =& P^b_{i_R, j_R} + \sum_{1 \leq k_1 \leq k_2 \leq L_S} P^{Is}_{i_R, j_R, k_1, k_2} + P^{Ie}_{i_R, j_R, k_1, k_2}, \\
P(i_S, j_S) =& P^b_{i_S, j_S} + \sum_{1 \leq k_1 \leq k_2 \leq L_R} P^{Is'}_{k_1, k_2, i_S, j_S} + P^{Ie}_{k_1, k_2, i_S, j_S}, \\
\begin{split} \label{equ-bprob}
P(i_R, j_S)=& \sum_{i_R < k_1 \leq L_R \atop 1 \leq k_2 < j_S} P^{Ih}_{i_R, k_1, k_2, j_S} + 
\sum_{1 \leq k_1 \leq i_R < k_3 \leq L_R \atop 1 \leq k_2 < j_S \leq k_4 \leq L_S}
P^{Ihh}_{k_1, k_3, k_2, k_4} \frac{Q^{Ih}_{k_1, i_R, j_S, k_4} Q_{i_R+1, k_3} Q_{k_2, j_S-1}} 
{Q^{Ihh}_{k_1, k_3, k_2, k_4}} + \\
& \sum_{1 \leq k_1 \leq i_R < k_3 \leq L_R \atop 1 \leq k_2 < j_S \leq k_4 \leq L_S}
P^{Ihb}_{k_1, k_3, k_2, k_4} \frac{Q^{Ih}_{k_1, i_R, j_S, k_4} (Q^{bz}_{i_R+1, k_3} Q_{k_2, j_S-1} +
Q^{bz}_{k_2, j_S-1})} {Q^{Ihb}_{k_1, k_3, k_2, k_4}}.
\end{split}
\end{align}
Equation (\ref{equ-bprob}) consists of two types of terms: 1) the probability that $i_R \circ j_S$ is on the left of a $Q^{Ih}$ component, which includes the cases where $i_R \circ j_S$ is in the middle of a hybrid component (see Figure \ref{fig-ih}), and 2) the probability that $i_R \circ j_S$ is on the right of a $Q^{Ih}$ component.

\subsection{Sampling algorithm}
In this section, we present an efficient algorithm to generate random samples 
from the Boltzmann ensemble of interaction structures. Each structure is drawn with probability equal to its Boltzmann probability. Let $n = L_R$ and $m = L_S$.
A na\"ive sampling algorithm, similar to the Ding-Lawrence algorithm \cite{DinLaw03}, has $O(n^2 m^2)$ time complexity in our case. In this paper, we give an efficient algorithm, which is inspired by Ponty's boustrophedon method \cite{Ponty08}, to improve the time complexity to $O((n + m)^2\log(n + m))$. 

\begin{figure}[h]
\begin{center}
\begin{tabular}{cc}
\begin{picture}(0,0)%
\includegraphics{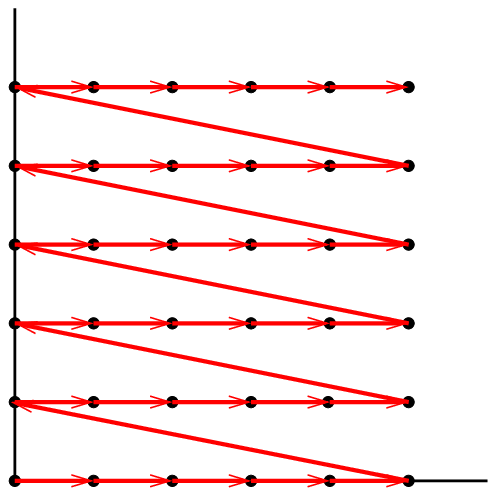}%
\end{picture}%
\setlength{\unitlength}{1657sp}%
\begingroup\makeatletter\ifx\SetFigFontNFSS\undefined%
\gdef\SetFigFontNFSS#1#2#3#4#5{%
  \reset@font\fontsize{#1}{#2pt}%
  \fontfamily{#3}\fontseries{#4}\fontshape{#5}%
  \selectfont}%
\fi\endgroup%
\begin{picture}(6528,6082)(1156,-5642)
\put(1531,-2401){\makebox(0,0)[lb]{\smash{{\SetFigFontNFSS{11}{13.2}{\rmdefault}{\mddefault}{\updefault}{\color[rgb]{0,0,0}$\cdot$}%
}}}}
\put(1531,-2086){\makebox(0,0)[lb]{\smash{{\SetFigFontNFSS{11}{13.2}{\rmdefault}{\mddefault}{\updefault}{\color[rgb]{0,0,0}$\cdot$}%
}}}}
\put(1531,-1771){\makebox(0,0)[lb]{\smash{{\SetFigFontNFSS{11}{13.2}{\rmdefault}{\mddefault}{\updefault}{\color[rgb]{0,0,0}$\cdot$}%
}}}}
\put(7156,-4876){\makebox(0,0)[lb]{\smash{{\SetFigFontNFSS{12}{14.4}{\rmdefault}{\mddefault}{\updefault}{\color[rgb]{0,0,0}$k_1$}%
}}}}
\put(1486, 29){\makebox(0,0)[lb]{\smash{{\SetFigFontNFSS{12}{14.4}{\rmdefault}{\mddefault}{\updefault}{\color[rgb]{0,0,0}$k_2$}%
}}}}
\put(1936,-5461){\makebox(0,0)[lb]{\smash{{\SetFigFontNFSS{11}{13.2}{\rmdefault}{\mddefault}{\updefault}{\color[rgb]{0,0,0}$i_R$}%
}}}}
\put(1576,-5101){\makebox(0,0)[lb]{\smash{{\SetFigFontNFSS{11}{13.2}{\rmdefault}{\mddefault}{\updefault}{\color[rgb]{0,0,0}$i_S$}%
}}}}
\put(1171,-4246){\makebox(0,0)[lb]{\smash{{\SetFigFontNFSS{10}{12.0}{\rmdefault}{\mddefault}{\updefault}{\color[rgb]{0,0,0}$i_S+1$}%
}}}}
\put(1171,-3346){\makebox(0,0)[lb]{\smash{{\SetFigFontNFSS{10}{12.0}{\rmdefault}{\mddefault}{\updefault}{\color[rgb]{0,0,0}$i_S+2$}%
}}}}
\put(1486,-601){\makebox(0,0)[lb]{\smash{{\SetFigFontNFSS{11}{13.2}{\rmdefault}{\mddefault}{\updefault}{\color[rgb]{0,0,0}$j_S$}%
}}}}
\put(5041,-5506){\makebox(0,0)[lb]{\smash{{\SetFigFontNFSS{11}{13.2}{\rmdefault}{\mddefault}{\updefault}{\color[rgb]{0,0,0}$\cdots$}%
}}}}
\put(6346,-5461){\makebox(0,0)[lb]{\smash{{\SetFigFontNFSS{11}{13.2}{\rmdefault}{\mddefault}{\updefault}{\color[rgb]{0,0,0}$j_R$}%
}}}}
\put(2566,-5461){\makebox(0,0)[lb]{\smash{{\SetFigFontNFSS{10}{12.0}{\rmdefault}{\mddefault}{\updefault}{\color[rgb]{0,0,0}$i_R+1$}%
}}}}
\put(3601,-5461){\makebox(0,0)[lb]{\smash{{\SetFigFontNFSS{10}{12.0}{\rmdefault}{\mddefault}{\updefault}{\color[rgb]{0,0,0}$i_R+2$}%
}}}}
\end{picture}%
& 
\begin{picture}(0,0)%
\includegraphics{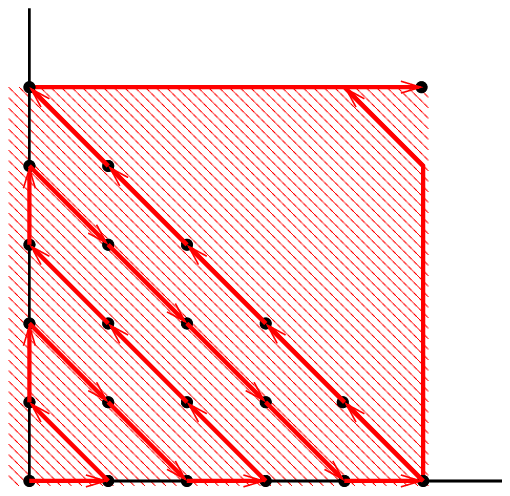}%
\end{picture}%
\setlength{\unitlength}{1657sp}%
\begingroup\makeatletter\ifx\SetFigFontNFSS\undefined%
\gdef\SetFigFontNFSS#1#2#3#4#5{%
  \reset@font\fontsize{#1}{#2pt}%
  \fontfamily{#3}\fontseries{#4}\fontshape{#5}%
  \selectfont}%
\fi\endgroup%
\begin{picture}(6888,6150)(796,-5710)
\put(7156,-4876){\makebox(0,0)[lb]{\smash{{\SetFigFontNFSS{12}{14.4}{\rmdefault}{\mddefault}{\updefault}{\color[rgb]{0,0,0}$k_1$}%
}}}}
\put(1486, 29){\makebox(0,0)[lb]{\smash{{\SetFigFontNFSS{12}{14.4}{\rmdefault}{\mddefault}{\updefault}{\color[rgb]{0,0,0}$k_2$}%
}}}}
\put(5671,-5506){\makebox(0,0)[lb]{\smash{{\SetFigFontNFSS{11}{13.2}{\rmdefault}{\mddefault}{\updefault}{\color[rgb]{0,0,0}$\cdots$}%
}}}}
\put(6211,-5551){\makebox(0,0)[lb]{\smash{{\SetFigFontNFSS{11}{13.2}{\rmdefault}{\mddefault}{\updefault}{\color[rgb]{0,0,0}$\left[ \frac{i_R+j_R}{2} \right]$}%
}}}}
\put(4366,-5506){\makebox(0,0)[lb]{\smash{{\SetFigFontNFSS{10}{12.0}{\rmdefault}{\mddefault}{\updefault}{\color[rgb]{0,0,0}$j_R-1$}%
}}}}
\put(3376,-5551){\makebox(0,0)[lb]{\smash{{\SetFigFontNFSS{10}{12.0}{\rmdefault}{\mddefault}{\updefault}{\color[rgb]{0,0,0}$i_R+1$}%
}}}}
\put(2836,-5461){\makebox(0,0)[lb]{\smash{{\SetFigFontNFSS{11}{13.2}{\rmdefault}{\mddefault}{\updefault}{\color[rgb]{0,0,0}$j_R$}%
}}}}
\put(1936,-5461){\makebox(0,0)[lb]{\smash{{\SetFigFontNFSS{11}{13.2}{\rmdefault}{\mddefault}{\updefault}{\color[rgb]{0,0,0}$i_R$}%
}}}}
\put(1576,-5101){\makebox(0,0)[lb]{\smash{{\SetFigFontNFSS{11}{13.2}{\rmdefault}{\mddefault}{\updefault}{\color[rgb]{0,0,0}$i_S$}%
}}}}
\put(1576,-4201){\makebox(0,0)[lb]{\smash{{\SetFigFontNFSS{11}{13.2}{\rmdefault}{\mddefault}{\updefault}{\color[rgb]{0,0,0}$j_S$}%
}}}}
\put(811,-646){\makebox(0,0)[lb]{\smash{{\SetFigFontNFSS{11}{13.2}{\rmdefault}{\mddefault}{\updefault}{\color[rgb]{0,0,0}$\left[ \frac{i_S+j_S}{2} \right]$}%
}}}}
\put(1621,-1951){\makebox(0,0)[lb]{\smash{{\SetFigFontNFSS{11}{13.2}{\rmdefault}{\mddefault}{\updefault}{\color[rgb]{0,0,0}$\cdot$}%
}}}}
\put(1621,-1636){\makebox(0,0)[lb]{\smash{{\SetFigFontNFSS{11}{13.2}{\rmdefault}{\mddefault}{\updefault}{\color[rgb]{0,0,0}$\cdot$}%
}}}}
\put(1621,-1321){\makebox(0,0)[lb]{\smash{{\SetFigFontNFSS{11}{13.2}{\rmdefault}{\mddefault}{\updefault}{\color[rgb]{0,0,0}$\cdot$}%
}}}}
\put(1126,-2446){\makebox(0,0)[lb]{\smash{{\SetFigFontNFSS{10}{12.0}{\rmdefault}{\mddefault}{\updefault}{\color[rgb]{0,0,0}$j_S-1$}%
}}}}
\put(1171,-3346){\makebox(0,0)[lb]{\smash{{\SetFigFontNFSS{10}{12.0}{\rmdefault}{\mddefault}{\updefault}{\color[rgb]{0,0,0}$i_S+1$}%
}}}}
\end{picture}%
\\
(a) & (b)
\end{tabular}
\caption{(a) Na\"ive traversal of indices, (b) Balanced traversal of indices.\label{fig-indices}}
\end{center}
\end{figure}

Our algorithm is iterative conditioning-sampling, based on the Ding-Lawrence algorithm. It starts with $Q^{I}_{1, n, 1, m}$ on top of an empty stack. In each step, our algorithm pops the top of the stack, which is a partition function term such as $Q^{Ia}_{i_R, j_R, i_S, j_S}$. It selects a recursion case, such as
the last case (rightmost) in Figure \ref{fig-ia} and samples a pair of indices $k^*_1, k^*_2$ (or a single index in the case of a single-strand partition function) with appropriate probability. For this example, the probability of indices $k_1 \in (i_R, j_R], k_2 \in [i_S, j_S)$ in the last case of $Q^{Ia}$ is
\begin{equation} \label{equ:pi}
\pi(k_1, k_2) = \frac{Q^{Ie}_{i_R, k_1, k_2, j_S} Q^{I}_{k_1+1, j_R, i_S, k_2-1}}{
\sum_{i_R < k_1 \leq j_R \atop i_S \leq k_2 < j_S} Q^{Ie}_{i_R, k_1, k_2, j_S} Q^{I}_{k_1+1, j_R, i_S, k_2-1}}.
\end{equation}
Let $\pi(i_R, \cdot) = \pi(\cdot, j_S) = 0$.
To describe the na\"ive approach first let 
\begin{equation} \label{equ-psi}
\psi(v) = \sum_{0 \leq t < v} \pi(i_R + (t \mbox{ mod } (j_R - i_R + 1)),
i_S + \left[\frac{t}{j_R - i_R + 1}\right]). 
\end{equation}
Figure \ref{fig-indices}(a) shows how the two-dimensional array of indices is traversed in $\psi$.
Note that (\ref{equ:pi}) and (\ref{equ-psi}) imply that $\psi((j_R - i_R + 1)(j_S - i_S + 1)) = 1$.
To properly sample $k^*_1, k^*_2$, our algorithm first generates a uniform random number $\alpha^* \in [0, 1+\pi(j_R, j_S))$. Let $v^*$ be such that $\psi(v^*) \leq \alpha^* < \psi(v^*+1)$, and let
\begin{align} \label{equ-index}
k_1^* &= i_R + (v^* \mbox{ mod } (j_R - i_R + 1)), \\
k_2^* &= i_S + \left[\frac{v^*}{j_R - i_R + 1}\right]. 
\end{align}
It is clear that $k^*_1, k^*_2$ are sampled according to $\pi$ distribution in (\ref{equ:p}).
Finally, $Q^{Ie}_{i_R, k^*_1, k^*_2, j_S}$ and $Q^{I}_{k^*_1+1, j_R, i_S, k^*_2-1}$ are pushed onto the stack. The algorithm terminates whenever the stack is empty. No matter in which order the indices $k_1, k_2$ are inspected in $\psi$, it takes $O(n m)$ time in the worst case to determine $v^*$. Therefore, the worst case running time of this na\"ive algorithm for a single sample structure is  $O(n^2 m^2)$. 

Our algorithm's speed-up comes from the following trick: let the worst case correspond to $k^*_1 = [(i_R+j_R)/2]$ and $k^*_2 = [(i_S+j_S)/2]$. Using this trick, the problem is split in an (almost) balanced way in every step. More precisely, our algorithm uses the traversal scheme of Figure \ref{fig-indices}(b) in $\psi$ instead of the scheme of Figure \ref{fig-indices}(a). In that case, the following lemma
characterizes the cost of each step $c(k^*_1, k^*_2)$ in the algorithm.

{\lemma
The cost
of sampling $k^*_1, k^*_2$ in our scheme shown in Figure \ref{fig-indices}(b) satisfies
\begin{equation}
c(k^*_1, k^*_2) \leq 2\left(\min(k^*_1 - i_R, j_R - k^*_1) + \min(k^*_2 - i_S, j_S - k^*_2) + 2\right)^2.
\end{equation}\label{lem-cost}}

\paragraph{Complexity analysis} Let $f(n, m)$ denote the worst case running time of our sampling algorithm for two nucleic acids of length $n$ and $m$. In that case, $f$ satisfies the following recursive inequality
\begin{equation}\label{equ-inequ}
f(n, m) \leq 2 f(\frac{n}{2}, \frac{m}{2}) + \frac{\left(n + m + 4\right)^2}{2}.
\end{equation}
It follows from (\ref{equ-inequ}) that $f(n, m)$ is $O((n+m)^2 \log(n+m))$. Hence, the following theorem holds:
\begin{theorem}
The 
worst case time complexity of our algorithm is $O((n+m)^2 \log(n + m))$.
\end{theorem}

\section{Results}
We implemented the centroid, base pair probabilities, and our sampling algorithms in the new version of {\tt piRNA} which is implemented in C++ and is parallelized with OpenMP. Our experiments were run on an IBM shared memory machine with 64 PPC CPUs and
256GB of RAM. Using {\tt piRNA}, we predicted the centroid for five interacting RNA pairs in Table \ref{tab-results}. The longest experiment corresponds to OxyS-fhlA pair which took about 4 days. We used the exact centroid computed using the base pair probabilities in this study. In future work, we would like to explore Ding et al. approach which consists of sampling the ensemble and clustering samples. Centroids of the clusters are used as candidate structures instead of
the exact centroid of the ensemble.

\begin{table}[h]\label{tab-results}
\begin{center}
\begin{tabular}{|c|c|c|c|c|c|c|c|}
\hline
RNA pairs & \multicolumn{3}{c|}{Sensitivity} & \multicolumn{3}{c|}{PPV} & Reference \\
 & \multicolumn{1}{c}{\tt piRNA} & \multicolumn{1}{c}{\tt inteRNA} & \multicolumn{1}{c|}{Kato et al.} & 
 \multicolumn{1}{c}{\tt piRNA} & \multicolumn{1}{c}{\tt inteRNA} & \multicolumn{1}{c|}{Kato et al.} & \\
\hline
Tar-Tar* & 1.0 & 1.0 & 1.0 & 0.875 & 0.875 & 0.933 & \cite{Chang97} \\
R1inv-R2inv & 0.900 & 1.0 & 0.900 & 0.900 & 1.0 & 0.947 & \cite{Rist01}\\
DIS-DIS & 1.0 & 0.785 & 0.785 & 1.0 & 0.785 & 0.785 & \cite{Paillart96}\\
CopA-CopT & 1.0 & 0.863 & 0.909 & 1.0 & 0.760 & 0.800 & \cite{Malmgren97} \\
OxyS-fhlA & 0.714 & - & - & 0.746 & - & - & \cite{ArgAlt00}\\
\hline
Average & {\bf 0.922} & 0.912 & 0.898 & {\bf 0.904} & 0.855 & 0.866 & \\
\hline
\end{tabular}
\caption{Comparison of the sensitivity and PPV of RNA-RNA interaction structure prediction by {\tt piRNA} centroid prediction with those of {\tt inteRNA} \cite{AlkKarNadSahZha06} and Kato et al. software \cite{Kato09}.}
\end{center}
\end{table}

Table \ref{tab-results} summarizes the specificity and positive predictive value (PPV) of our RNA-RNA interaction structure prediction by centroid prediction. We considered Kato et al. dataset \cite{Kato09} excluding RepZ-IncRNA$_{54}$ and including OxyS-fhlA. Due to limitation on the computational resources, we replaced RepZ-IncRNA$_{54}$ with OxyS-fhlA as RepZ-IncRNA$_{54}$ exhibits a CopA-CopT-like  secondary structure whereas OxyS-fhlA has a different structure with two kissing hairpins. 
We expected centroid prediction
to outperform minimum-free-energy prediction, and our expectation is verified. 

\section{Conclusions and future work}
We presented base pair probabilities of interacting nucleic acids based on our previous interaction partition function algorithm {\tt piRNA} \cite{Chitsaz09}. The centroid of the Boltzmann ensemble is
computed from the base pair probabilities.  
We also presented an efficient algorithm to sample interaction structures from the ensemble. Our sampling algorithm uses a balanced scheme for traversing indices (depicted in Figure \ref{fig-indices}(b)). The worst case running time complexity of our algorithm is $O((n+m)^2\log(n+m))$, in which $n$ and $m$ are the lengths of input strands.
These algorithms are incorporated in the new version of {\tt piRNA}.

In future work, we would like to explore Ding et al. approach which consists of sampling the ensemble and clustering samples. Centroids of the clusters are used as candidate structures instead of
the exact centroid of the ensemble. We believe success of such an approach critically depends on the clustering method, therefore, we would like to study sampling-clustering algorithms in future work.

\paragraph*{Acknowledgement}
H. Chitsaz received funding from Combating Infectious Diseases (BCID) initiative.
S.C. Sahinalp was supported by Michael Smith Foundation for Health Research Career Award.

\end{document}